\newcommand{\diracslash}[1]{#1\llap{/\kern2pt}}

\newcommand{\be}{\begin{equation}}
\newcommand{\ee}{\end{equation}}
\newcommand{\bea}{\begin{eqnarray}}\index{\footnote{}}
\newcommand{\eea}{\end{eqnarray}}
\newcommand{\ba}[1]{\begin{array}{#1}}
\newcommand{\ea}{\end{array}}

\documentclass[12pt]{iopart}
\usepackage{graphicx}
\begin{document}
\setlength{\topmargin}{0.2in}

\title[Suppresion of power-broadening]{Suppression of power-broadening in strong-coupling photoassociation in the presence of a
Feshbach resonance}
\author{Bimalendu Deb$^{1,2}$ and Arpita Rakshit$^1$}
\address{$^1$Department of Materials Science, and $^2$Raman Center for
Atomic, Molecular and Optical Sciences, Indian Association for the
Cultivation of Science (IACS), Jadavpur, Kolkata 700032. INDIA}

\begin{abstract}

Photoassociation (PA) spectrum in the presence of a magnetic Feshbach resonance is analyzed.
Nonperturbative solution of the problem yields analytical expressions for PA linewidth and shift
which are applicable for  arbitrary PA laser intensity and magnetic field tuning of Feshbach
Resonance. We show that by tuning magnetic field close to Fano minimum, it is possible to suppress
power broadening at increased laser intensities. This occurs due to quantum interference of PA
transitions from unperturbed and perturbed continuum. Line narrowing at high laser intensities is
accompanied by large spectral shifts. We briefly discuss important consequences of line
narrowing in cold collisions.

\end{abstract}

\pacs{34.50.Cx, 34.80.Dp, 32.70.Jz, 34.80.Pa}

\maketitle

\section{Introduction}
Over the years, photoassociation and Feshbach resonance have become important tools in manipulation
of ultracold collisions. Recent experimental \cite{Junker:prl:2008, Winkler, ni} and theoretical
\cite{Mackie:prl:2008, pellegrini:njp:2009,cote2,Kuznetsova} works on photoassociation (PA) near a
magnetic field Feshbach resonance (MFR) \cite{mfr} give rise to the exciting possibilities of
coherent control of atom-molecule conversion. In a recent experiment, Junker {\it et al.}
\cite{Junker:prl:2008} have demonstrated asymmetric spectral line shape and saturation in PA in the
presence of  MFR. Asymmetric line shape is a characteristic feature of Fano effect \cite{fano} which
arises in different areas such as atomic \cite{madden}, particle \cite{muller} and condensed matter
physics \cite{faist}. When a pair of colliding atoms under the influence of Feshbach resonance are photoassociated
into  an excited molecular state, PA transitions can occur in two competing pathways. The presence of
a Feshbach resonance largely perturbs the continuum states of colliding atoms. As a result, continuum
states get hybridised with one or more bound states embedded in the continuum. PA transitions from unperturbed and perturbed continuum states can interfere resulting in asymmetric Fano line shape. The unique aspect of MFR induced Fano effect in PA is that the continuum-bound coherence can be controlled by tuning the magnetic field. Feshbach resonances and photoassociation have a common feature: Both are the effects of continuum-bound interacting systems. In the case of Feshbach resonance, an initial continuum state is
coupled to a bound state embedded in the continuum by hyperfine interaction of two atoms. In the case of PA, the continuum state gets coupled to an excited diatomic molecular state by a single photon. Both can be treated within the framework of Fano's theory. In
fact, based on Fano's method, theory of one- and two-photon PA have been developed by Bohn and Julienne \cite{semian, semian2}   Two-photon PA \cite{two-photon} has been shown to lead to quantum interference. Furthermore, quantum interference has been
demonstrated in coherent formation of molecules \cite{atom-molecule}.

Apart from MFR-induced modification, PA under intense laser fields (strong-coupling regime) can further modify
the continuum states due to strong continuum-bound dipole coupling.  Fano effect in the strong field
regime \cite{eberly,zoller,gsa:prl:1982} leads to ``confluence of bound-free coherences''
\cite{eberly} which can result in a number of effects such as line narrowing \cite{Neukammer},
Autler-Townes splitting \cite{eberly} and  nonlinear Fano effect \cite{kroner}. Continuum-bound coherences in strongly interacting ultracold atoms can lead a number of profound physical effects
which may not be observed in the weak coupling regime. A decade ago, Vuletic {\it et al.} \cite{vuletic}
experimentally observed line narrowing in the spectrum of trap loss of atoms due to a tunable
Feshbach resonance. In Fano's Theory \cite{fano}, a continuum interacting with a bound state is
exactly diagonalized leading to a ``dressed" continuum state. In linear Fano Theory, optical
transition matrix element between this dressed continuum and any other bound state is obtained
perturbatively by the use of Fermi Golden rule. In the strong coupling regime, optical coupling of
the dressed continuum with any other state needs to be treated nonperturbatively either by ``double
diagonalization" technique \cite{eberly} or by other diagonalization techniques such as used in
\cite{gsa}.

Here we explore the possibility of suppression of power-broadening in strong-coupling PA  by
manipulation of continuum-bound coherences with Feshbach resonance. As the simplest possible model,
we consider optically coupled two bound states interacting with a common continuum as shown in figure 1. We demonstrate that by tuning the
magnetic field close to Fano minimum where excitation probability vanishes \cite{fano}, it is possible to obtain line narrowing in PA
spectrum with large shifts at high laser intensities.  Large light-shift with narrow linewidth may be
useful in efficient tuning of elastic scattering length by optical means \cite{fedichev,fatemi, deb}.

 The paper is organised in the following way. In the following section, we discuss in brief the formulation of the problem. In section 3, we present the analytical results on line narrowing and enhencement of shifts. In section 4, we discuss the numerical results and the paper is concluded in section 5.

\begin{figure}
 \includegraphics[width=6.00 in]{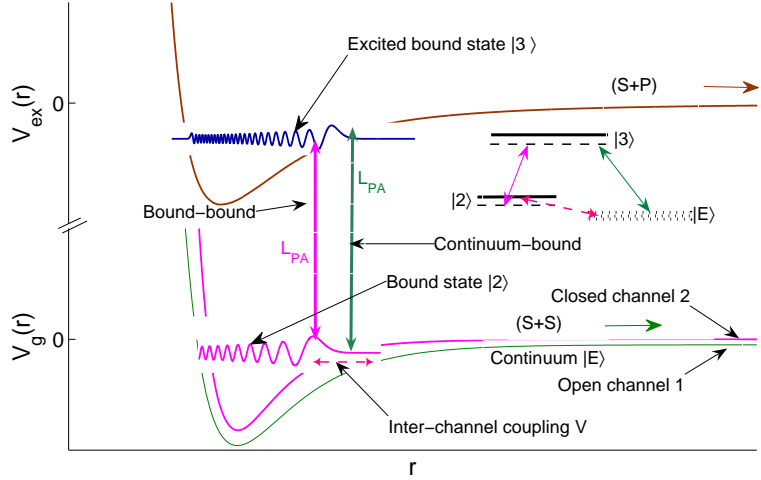}
 % figfannew1.eps: 1179668x1179666 pixel, 300dpi, 9987.86x9987.84 cm, bb=   51   184   558   607
 \caption{A schematic diagram showing the coupling between the bound state (magenta) $\mid 2 \rangle$
  and and continuum (green) $\mid E \rangle$ with the excited bound state (blue) $\mid 3 \rangle$ via same
  laser $L_{PA}$( magenta and green double-arrow vertical lines). Red dashed line indicates the hyperfine coupling
   between the open and closed channel. }
 \label{Figure 1.}
\end{figure}
\section{Formulation of the problem}
The dressed state of a system of two colliding  atoms interacting
with a PA laser in the presence of MFR can be written as
\bea \mid \Psi_{E} \rangle = \frac{1}{r}\left[ \int d E' b_{E'} \Psi_{E'}(r) \mid 1 \rangle + \sum_{i=2,3} \Phi_i(r) \mid i \rangle \right]\label{ps}\eea
where $r$ is the relative coordinate of the two atoms, $E$ is an energy eigenvalue, $\mid 1 (2)  \rangle$
represents the internal electronic states of   open(closed)
channel $\mid 1(2) \rangle$ and $\mid 3 \rangle$ denotes the electronic state of the
excited molecule. $\Phi_{i}$ ($i=2,3$)  and $\Psi_{E'}$ are the wave functions of
perturbed bound and continuum states, respectively. Here $b_{E'}$ is density of states of the unperturbed
 continuum. In writing the above equation, we have assumed that PA laser can couple only a particular
 ro-vibrational state of the excited molecule; and the bound state $\Phi_2$ has zero angular momentum.
 Let the  hyperfine spin coupling between the channels 1 and 2 be denoted by $V(r)$.  Let $\Omega_1(r)$
 and $\Omega_2(r)$ represent the molecular Rabi couplings of the excited state $\mid 3 \rangle $
with the ground states  $\mid 1 \rangle$ and $\mid 2 \rangle $, respectively. In the absence of these
three couplings, let the unperturbed bound states be denoted by $\phi_3(r)$ and  $\phi_2(r)$ with bound
state energies $E_3$ and $E_2$ respectively; and the unperturbed continuum states by $\psi_{E'}$ with
asymptotic collision energy $E'$. With the use of these unperturbed solutions, we construct three
Green's functions $G_E (r,r')$, $G_2(r, r')$ and $G_3(r,r')$ which correspond to the channels 1, 2 and 3,
respectively.
The continuum Green's function $G_E(r,r')$ can be written as
$G_E (r,r') = - \pi\psi_E^{reg}(r_<)\psi_E^{+}(r_>)$, where $r_{<(>)}$ implies either $r$ or $r'$ whichever
 is smaller (greater) than the other. Here $\psi_E^{+}(r) = \psi_E^{irr} + i\psi_E^{reg}$ where $\psi_E^{reg} $
 and $\psi_E^{irr}$ represent regular and irregular scattering functions, respectively. Asymptotically,
  $ \psi_{E}^{0 , reg}(r) \sim
j_{0}\cos\eta_{0} - n_{0}\sin\eta_{0}$ and  $ \psi_{E}^{0,
irr}(r) \sim -(n_{0}\cos\eta_{0} + j_{0}\sin\eta_{0}) $, where
$j_{0}$ and $ n_{0}$ are the spherical Bessel and Neumann
functions for partial wave $\ell = 0$ (s-wave) and $\eta_{0}$ is the s-wave phase shift
in the absence of laser and magnetic field couplings. The other two Green's functions correspond to
bound state solutions and are of the form  $G_3(r,r') =  - \frac{1}{\hbar
\delta + E - E_{3}  }\phi_{3}(r)\phi_{3}(r')$ and $G_2(r,r') = - \frac{1}{E - E_{2}  }\phi_{2}(r)\phi_{2}(r')$.
 Here $\delta = \omega_L - \omega_A$ is the laser-atom detuning, the bound state energy
$E_3$ is measured from the excited state threshold $\hbar\omega_A$.   In order to include the
spontaneous emission of the excited state, following the prescription given by Bohn and Julienne
\cite{semian}, we introduce in our model an artificial  channel with channel state $\psi_{art}$ and its coupling $V_{art}$
with excited state $ \mid 3 \rangle $; and thereby identify spontaneous linewidth $\hbar
\gamma = 2\pi |\langle \psi_{art} \mid V_{art} \mid \phi_3 \rangle |^2$.

Next, let us introduce a low energy dimensionless interaction parameter
 \bea \beta(k)= -\cot \eta_{res}(k)\simeq  (k a_{res})^{-1} + \frac{1}{2} r_ek\label{bta} \eea
where $ \eta_{res}(k)$ is the Feshbach resonance phase
shift and $r_e$ is  related to Feshbach resonance linewidth $\Gamma_{r} \sim k C $ by $r_e =
2\hbar/(\mu C)$ where C is a constant and $\mu$ is the reduced mass of the two atoms.  $a_{res}$ is
related to the applied magnetic field $B$ by  $a_{res}= - \frac{ a_{bg} \Delta}{B - B_0}$
\cite{verhaar}, where $a_{bg}$ is the background scattering length, $B_0$ is the  resonance magnetic field, $\Delta$ is a parameter which depends
on $\Gamma_{r}$ and magnetic moments of the atoms at $B_{0}$. Having done so, we can obtain the
solutions in the form $\Phi_3(r) = A_{PA} \phi_3(r)$ and  $\Phi_2(r) = A_{CC} \phi_2(r)$. The
explicit form of $A_{PA}$  \cite{gsa} is given by \bea A_{PA}= \frac{\exp(i\eta_0) (\beta + q ) \pi
\Omega_{3 E} }{( \beta + i ) \{\Delta_p + i \hbar (\gamma +  \Gamma_{PA})/2\} - \hbar \Gamma_{PA} (q - i)^2 / 2}
\label{apa}\eea where $\Delta_p = \hbar\delta + E - (E_3 + E_{shift}^0) $, $E_{shift}^0$ is the
energy shift in the absence of Feshbach resonance, $\Omega_{3 E} = \int d r \phi_3(r) \Omega_1(r)
\psi_E^{reg}$ is the continuum-bound molecular dipole coupling and $\Gamma_{PA}$ is the stimulated
linewidth of PA given by $ \Gamma_{PA} = 2\pi|\Omega_{3E}|^2$. Here $q$  is Fano's asymmetry
parameter defined by
 \bea
 q =\frac{\Omega_{32} + V_{32} }{ \pi \Omega_{3E} {V}_{2E}}\eea
 where $\Omega_{3 2}=\int dr \phi_3(r) \Omega_2(r) \phi_2$ is the bound-bound Rabi coupling and $V_{23}= \int\int{dr' dr \phi_2(r)
V(r) Re[G_E(r,r')] \Omega(r') \phi_3(r')}$ represents an effective interaction between the two bound
states mediated through their couplings with the common continuum. Note that $q$ is independent of
laser intensity. In the limit $k \rightarrow 0$, both $\Omega_{32}$ and  $V_{32}$ become
energy-independent while  $\Omega_{3E} {V}_{2E} \sim k$. Thus at low energy, $q \sim 1/k$. Detailed
derivation of the dressed state of \label{s} is given elsewhere \cite{gsa}, but for completeness we reproduce the
derivation in Appendix A.

\section{Analytical results}
 The loss of atoms due to decay of excited state into the decay channels (which is modeled as an artificial scattering channel) is described by the PA rate
\bea K_{PA} = < | \langle \psi_{art} \mid V_{art} \mid \psi \rangle|^2 > \nonumber\\
= \frac{1}{hQ_T} \int{ dE_k  \frac{\hbar^2 \gamma \Gamma \exp( -E_k/K_B T ) }{(E_k - \Delta
E+\hbar\delta_p)^2+ \hbar^2(\gamma+\Gamma)^2/4}}, \eea where $\langle \cdots \rangle $ means thermal
averaging over the collision energy and $\hbar\delta_p = \hbar\delta - (E_3+E^0_{shift} -E_{th})$ is
the detuning parameter. Here $E_{th}$ is the threshold of the open channel. Here $Q_T=(2\pi\mu K_B T/h^2)$, $E_k =E-E_{th}$
and $K_B$ is the Boltzmann constant. $\Gamma$ is the linewidth (in the presence of MFR) given by
.\begin{figure}
 \includegraphics[width=4.75 in]{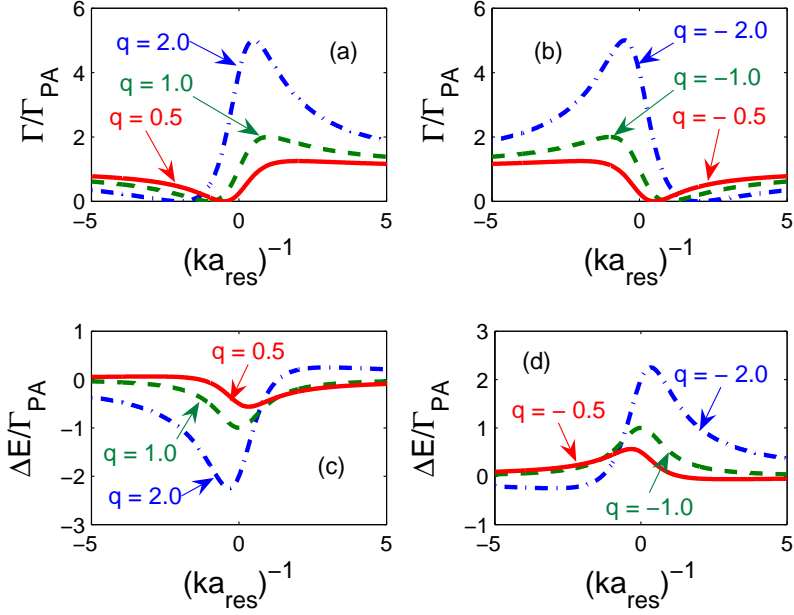}
 % untitled1.eps: 0x0 pixel, 300dpi, 0.00x0.00 cm, bb=   96   238   515   553
 \caption{Subplots (a) and (b) show $\Gamma/\Gamma_{PA}$ Vs $(ka_{res})^{-1}$ and subplots (c) and (d) exihibit
 $\Delta E/ \Gamma_{PA}$ Vs $(ka_{res})^{-1}$. Plots are for different values $q$ as indicated in the figure. $k$ is defined as $\frac{\hbar^2k^2}{2m}= K_BT$}
 \label{Figure 2.}
\end{figure}
\begin{eqnarray}
\Gamma = f(q, \beta) \Gamma_{p} = \frac{(\beta + q )^2}{\beta^2 + 1} \Gamma_{PA} \label{gamma}.
\end{eqnarray}
The extra shift caused by MFR is given by
\begin{eqnarray}
\Delta{E} = \frac{1}{2}\left[ \frac{(q^2-1)\beta - 2 q}{\beta^2 + 1} \right] \Gamma_{PA} \label{dshift}.
\end{eqnarray}
\begin{figure}
 \includegraphics[width = 4.75 in]{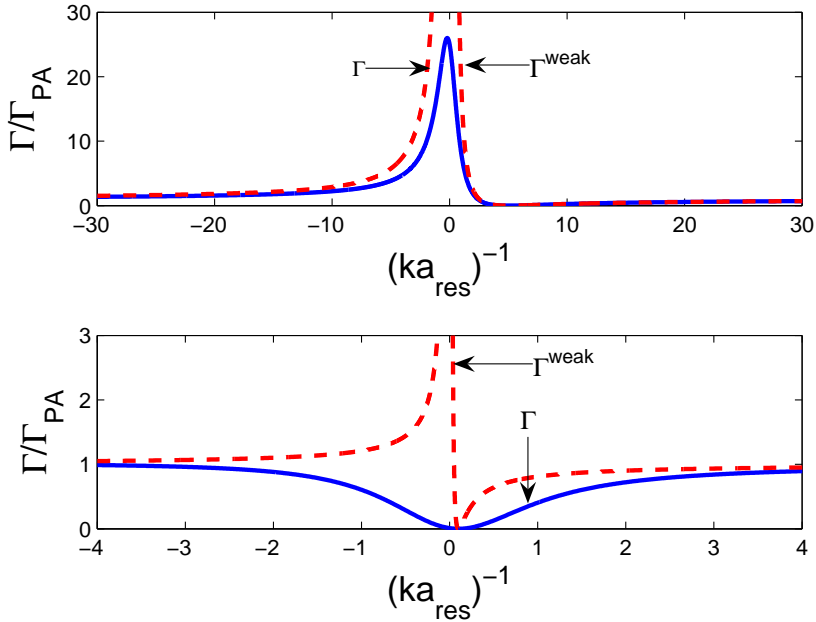}
  \caption{ Solid lines represent the linewidth $\Gamma$ (in unit of $\Gamma_{PA}$) given by the expression
  (\ref{gamma}) as a function of
   $(k a_{res})^{-1}$
  for $q = -5.0$ (upper panel) and $q = -0.1$
  (lower panel). Dotted lines represent
 the low coupling expression of (\ref{cg}) for $C_{1} = 6$ and $C_2 = 1$ (upper panel); and
  $C_{1} = 0.1$ and $C_2 = 0$ (lower panel). The values of $C_1$ and $C_2$ are so chosen such that
  in the limit $a_{res} \rightarrow 0$, $\Gamma \rightarrow \Gamma^{weak}$.}
 \label{Figure 3.}
\end{figure}
(\ref{gamma}) shows that $\Gamma$ depends on a nonlinear function ($f(q, \beta)$) of $q$
and $\beta$.  Note that when $ \beta \rightarrow \pm \infty $, that is, far away from MFR,  $\Gamma
\rightarrow \Gamma_{PA}$. It is to be further noted that when $\beta = - q$, we have $\Gamma = 0$ and
$A_{PA} = 0$ at which PA ceases to occur. The Fano minimum is given by $\beta_{min} = - q$ or equivalently the corresponding magnetic field $B_{min}$. Thus $\Gamma$ can be made arbitrarily small by tuning
$\beta$ close to $-q$. It is possible to suppress power-broadening at increased laser intensities by
tuning the magnetic field $B$ close to  $B_{min}$.

Next, we discuss the weak-coupling limit of (\ref{gamma}) and (\ref{dshift}) when laser intensity is low. For this we first find the dressed continuum state in the limit $\Gamma_{PA}\rightarrow 0$. As $\Gamma_{PA} \rightarrow 0$,
$A_{PA} \rightarrow 0$, $A_{CC}$ can be expressed as \bea A_{CC} = - \sqrt{\frac{2}{\pi \Gamma_r}}
 \exp{i(\eta_0 + \eta_{res})} \sin{\eta_{res}}\eea
and $\psi_E(r)$ becomes \bea \psi_E(r) = \exp{i(\eta_0 + \eta_{res})}[\psi_E^{0,reg} \cos{\eta_{res}}
 + \psi_E^{0,irr} \sin{\eta_{res}}] \eea
 So in the limit of  $\Gamma_{PA} \rightarrow 0 $, at low energy the  state $\mid \Psi_E \rangle $  of (\ref{ps}) reduces to
 \bea \mid \Psi_E \rangle_0 =\frac{1}{r} [A_{CC}  \phi_2 (r) \mid 2 \rangle + \int b_{E'}  \psi_{E'}(r) dE'  \mid 1 \rangle]. \eea Taking $b_{E}' = \delta(E-E')$, the stimulated linewidth $ \Gamma^{weak}$ in the weak coupling limit is given by the Fermi golden rule expression \bea \Gamma^{weak} &=& 2\pi |\int r \phi_{VJ}(r) \langle 3 \mid \Omega_1(r) \mid \Psi_E \rangle_0 dr|^2 \nonumber\\
&=& \Gamma_{PA}  |1 + C_1 \sin{\eta_{res}} + C_2 \tan{\eta_{res}}|^2 \label{cg}\eea
 where $C_1 = \Omega_{3E}^{irr}/\Omega_{3E}$  and $C_2 = (-\sqrt{\frac{2}{\pi \Gamma_r}} ){\Omega_{32}}/{\Omega_{3E}}$ . Here
$\Omega_{3E}^{irr} = \langle 3 \mid \Omega_1 \mid \psi_E^{0,irr} \rangle $. The expression
$(\ref{cg})$ is in agreement with  (6) of  \cite{pellegrini:njp:2009}. When
$\eta_{res}\rightarrow \pi/2$, $\Gamma^{weak}$ diverges and hence $(\ref{cg})$ is not
valid near $\eta_{res} = \pi/2 $. In other words, $(\ref{cg})$ is not applicable close to
Feshbach resonance.

Here we show how the exact expressions of $\Gamma$ and $\Delta E$ as given by (\ref{gamma}) and (\ref{dshift}), respectively, do enable us to realize the possibility of suppression of power-broadening with enhenced light shift at increased laser intensities. It is clear from  (\ref{dshift}) that $\Delta E$ goes to zero as $\beta \rightarrow \pm \infty$. From (\ref{gamma}) and (\ref{dshift}), we notice that if laser intensity $I$ is increased by a factor $M_{I}$, to suppress
power-broadening $\beta$ is to be changed to $\beta'$ such that  $f(q, \beta') = M_I^{-1}f(q,
\beta)$. For $\beta \simeq -q$ we have $\Delta{E} \simeq  -q \Gamma_{PA}/2$ which is proportional to
$I$. Hence when line broadening is suppressed by the tunability of MFR, the total shift
$E_{shift} = E^0_{shift} + \Delta E$ remains proportional to $I$. The shift $E^0_{shift}$ is negative in the low energy regime. However, the total shift $E_{shift}$ in the presence of MFR can be positive or negative depending on the values of $q$
and $\beta$.

Finally we prove that line narrowing  in one-photon PA is not possible in the absence of coupling between open and closed channel.
It can be noticed that the PA laser can be tuned either near continuum-bound frequency in which
case $ \hbar\delta_p \simeq E_k $ or near bound-bound transition frequency $(
\tilde{E_3}-\tilde{E_2})/ \hbar$ in which case $-\hbar\delta_p = (\tilde{E_2}-E_{th}) = E_{res}  $.
Here $\tilde{E_3} = E_3 + E^0_{shift}$ and $\tilde{E_2} = E_2 + \Delta E_2$,  $\Delta E_2$ being the
shift of closed channel bound state due to its coupling V with the open channel. In the limit
$\Gamma_r \rightarrow 0$, \bea K_{PA} \simeq \frac{\hbar^2 \gamma \Gamma_{PA}}{ \left\lbrace
(E-E_3-E^0_{shift})+\hbar\delta- \frac{\hbar^2\Omega_{32}^2}{E-E_2}\right\rbrace^2 +i\hbar^2 (\gamma + \Gamma_{PA})^2/ 4}
\eea which is in agreement with the expression of $|S_{1g}|^2$ of  \cite{semian2} if we identify
$\Delta_1$and $\Delta_2$  of  \cite{semian2} with $ -\hbar\delta$ and $E_2$, respectively. In our
case there is only one laser coupling between the continuum and the excited bound state and also between the
two bound states. It is clear from the above expression that in the absence of coupling between  open
and closed channels the narrowing of PA linewidth is not be possible.

\section{Numerical results and discussion}

\begin{figure}
 \includegraphics[width = 4.75 in]{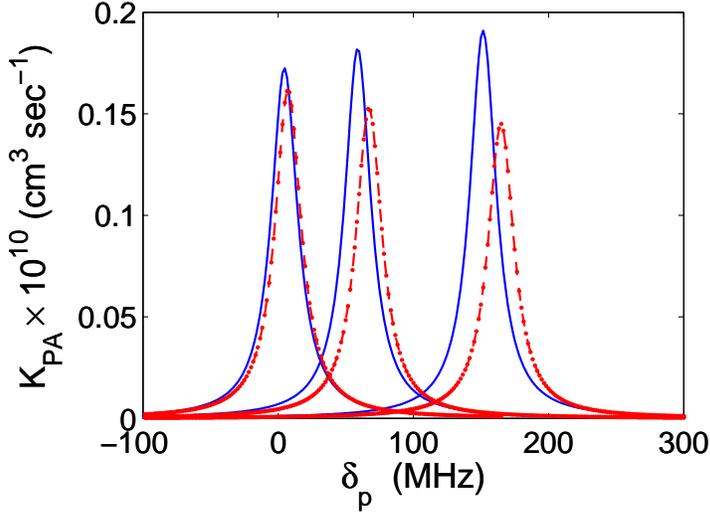}
  \caption{ $K_{PA}$ in cm$^3$ sec$^{-1}$ Vs.  detuning $\delta_p$ in MHz.  Each pair of dashed and solid
  curves are  obtained for $\Gamma_{PA} = 1.0$ MHz (left pair), $\Gamma_{PA} = 10.0$ MHz (middle pair) and
   $\Gamma_{PA} = 25.0$ MHz (right pair) for the fixed $q = -6.36$. For solid curves, magnetic fields are
   $B = 705.00$ G (left), $B = 708.56$ G (middle) and $B = 709.12$ G (right). For dash-dotted curves, these are
    $B = 713.19$ G (left), $B = 711.09$ G (middle) and $B = 710.68$ G (right). The magnetic fields are so
    chosen such that the linewidth $\Gamma$ remains fixed at 0.04 MHz.  }
 \label{Figure 4.}
\end{figure}
\begin{figure}
 \includegraphics[width = 5.75 in]{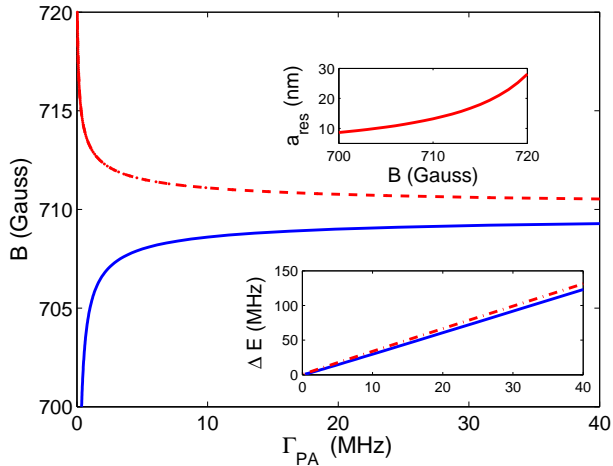}
\caption{The locus of $B$ and $\Gamma_{PA}$ for which the linewidth $\Gamma$ remains fixed at 0.04 MHz.
The lower inset shows the variation of $ \Delta{E} $ (in MHz) against  $\Gamma_{PA} $ (in MHz) and
the upper inset exhibits the variation $a_{res}$ (in nm) against $B$ (in Gauss) at the fixed $\Gamma = 0.04$ MHz.}
 \label{Figure 5.}
\end{figure}

 For numerical illustration, we consider a model system of two ground-state (S$_{1/2}$) $^7$ Li atoms undergoing PA from the ground
molecular configuration $ {^3}\Sigma^+_u$ to the vibrational state $v = 83$ of the excited molecular
configuration  1 ${^3}\Sigma_g^{+}$  which correlates asymptotically to 2S$_{1/2}$ + 2P$_{1/2}$ free
atoms \cite{prodan,abraham}.  All the relevant parameters $\gamma$, $E^0_{shift}$, $\Delta$, $ a_{bg} $  and $\Gamma_r$ are estimated from \cite{prodan}, \cite{Pollack} and \cite{chin}. In figure 2, we have plotted
$\Gamma/\Gamma_{PA}$ and $\Delta E/ \Gamma_{PA}$ against $1/ka_{res}$ for positive and negative q
values. The maximum and minimum values of linewidth would be observed for $ \beta = 1/q $ and $ \beta
= -q $, respectively. The magnitude of the change in shift due to PA in the presence of MFR is
significant near $\beta = -q$. Figure 3 clearly shows that the stimulated linewidth  in the weak coupling
limit, represented by dashed lines deviate appreciably from nonperturbative results as shown by solid lines. The deviations are the  most prominent in the region $ (k a_{res})^{-1} \simeq 0$ $(\eta_{res} \simeq \pi/2)$. Furthermore, for lower $q$ values  these two results deviate most significantly.  Figure 4 illustrates how to suppress power-broadening by  the appropriate tuning of magnetic field near $B_{min}$  and thereby to keep the total linewidth close
to the natural linewidth. There are two values of $\beta (B)$ and correspondingly two values of
$a_{res}$ where the linewidth $\Gamma$ can be kept fixed at a small value at an increased laser
intensity. In figure 5, we show  how to vary $\Gamma_{PA}$ (or laser intensity) and the magnetic
field in order to keep $\Gamma$ fixed at  0.04 MHz which is much smaller than the natural linewidth
$\gamma$ although $\Gamma_{PA}$ can be many orders of magnitude higher than $\gamma$. The lower inset
in figure 5 shows that the extra shift $\Delta E$  can exceed $\gamma$ by many orders of magnitude
while power-broadening is suppressed.

\section{Conclusion}

In conclusion, we have demonstrated  that linewidth of photoassociation spectrum can be narrowed down
close to the natural linewidth by making use of tunability of a magnetic Feshbach resonance.
Experimentally, the line narrowing effect may be inferred from PA spectra near $B_{min}$  at a high
intensity. This enhencement of the life time of excited molecular state may be beneficial for
population transfer from ground state collisional continuum to ground molecular state by two-photon
Raman-type PA. Furthermore, narrow linewidth with large shift will be useful for efficient
manipulation of scattering length by optical Feshbach resonance \cite{fedichev, fatemi}. This will be
particularly important for altering scattering amplitude of higher partial waves \cite{deb}.

\ack
One of us (AR) is grateful to CSIR, Govt. of India for a support.  BD is thankful and indebted to G. S. Agarwal for helpful discussions.

\appendix
\section*{Appendix-A}
\setcounter{section}{1}

The coupled differential equations for the model are given by \bea \left[
-\frac{\hbar^2}{2\mu}\frac{d^2}{d r^2}+B_J(r)\right] \Phi_3
&+& \left[ V_e(r)-\hbar\delta -E-i\hbar\gamma/2\right]  \Phi_3 \nonumber\\
&=& -\Omega_1\chi -\Omega_2\Phi_2,\eea\\
\bea \left[ -\frac{\hbar^2}{2\mu}\frac{d^2}{d r^2}+V_2(r)-E\right]
\Phi_2 = -\Omega_2^*\Phi_3 - V^* \chi, \eea\\
\bea \left[ -\frac{\hbar^2}{2\mu}\frac{d^2}{d r^2}+V_1(r)-E\right]\chi
=-\Omega_1^*\Phi_3 - V^*\Phi_2.\label{xi}\eea\\
Using Green's functions $G_3(r,r')$, we find $\Phi_3 = A_{PA}\phi_3 $, where \bea A_{PA} = \frac{\int dr'\left[\Omega_1(r')\chi(r') +
\Omega_2(r')\Phi_2(r')\right]}{\hbar\delta + E -E_3+i\hbar\gamma/2}\phi_3(r').\label{pa}\eea In a similar way, using $G_2(r,r')$ we can write $\Phi_2 = A_{CC} \phi_2$, where \bea A_{CC}=\frac{ [A_{PA}\Omega^*_{32}+\int dr' \chi(r') \Omega_1(r') \phi_2(r')]}
 {E-E_2}\eea
where $\Omega_{32} = \int d r \phi_{3}(r)
\Omega_{2}(r) \phi_{2}(r)$.
The continuum state can be written as $\chi(r) = \int dE' b_{E'}\psi_{E'}(r)$. Thus we obtain
\bea  - \frac{\hbar^2}{2\mu}\frac{d^2 }{d r^2} \psi_{E'}(r)
&+& [V_{1}(r) - E ]\psi_{E'}(r) = - \Omega_1^{*}(r)
\tilde{A}_{PA} \phi_{3}(r) \nonumber \\
 &-& \frac{( \tilde{V}_{2E} + \tilde{A}_{PA} \Omega_2) }{E- E_2} V(r) \phi_{3}(r)
\label{pseq}\eea
 where \bea \tilde{A}_{PA} = \frac{\tilde{\Omega}_{3E} (E - E_2) + \Omega_{32}
\tilde{V}_{2E}} {{\cal D} (E - E_{2}) - |\Omega_{32}|^2}.\label{tap} \eea
Here  ${\cal D } = \hbar \delta + E - E_{3} + i\hbar \gamma/2 $,
$\tilde{V}_{2E} = \int d r \phi_{2}(r) V(r) \psi_{E'}(r)$ and $
\tilde{\Omega}_{3E} = \int d r \phi_{3}(r) \Omega_{1}(r)
\psi_{E'}(r)$ . Using the Green's function $G_E (r,r')$, from equaion (\ref{pseq}) we obtain
\bea \psi_{E'} &=& \exp(i
\eta_0) \psi_{E'}^{0.reg} + \int d r' G_E(r, r') \left [
\Omega_1^{*}(r') \tilde{A}_{PA} \phi_{3}(r')
\right. \nonumber \\
 &+& \left. \frac{ \tilde{V}_{2E}^{*} +  \tilde{A}_{PA}
\Omega_{32 }}{(E - E_2)} V(r') \phi_{2}(r') \right ]. \label{si} \eea Now using this solution, we can
calculate the probability amplitude of excitation in the
following form \bea  \tilde{A}_{PA} &=&  \frac{\exp(i \eta_0) \left ( \Omega_{3E} + G
V_{2E}^{*} \right )} {{\cal D}  - \frac{|\Omega_{32}|^2}{E - E_2} - ( B_{p} + G B_f ) }
\label{apn} \eea where \bea V_{2E} = \int d r \phi_{2}(r) V(r) \psi_{E'}^{0,reg}(r)\eea and
\bea
\Omega_{3E} = \int d r \phi_{3}(r) \Omega_{1}(r) \psi_{E'}^{0,reg}(r). \eea The other
parameters here are \bea G = \frac{[\Omega_{32} +V_{32} - i\pi V_{2E}
\Omega_{3E}]}{[E - E_2 -\Delta {E_2} + i\pi |V_{2E}|^2]}\eea,
\bea B_p = E^0_{shift} - i\pi
|\Omega_{3E}|^2 - \Omega_{32 } \frac{[ -V_{32} + i\pi V_{2E}
\Omega_{3E}]}{(E- E_2)} \eea  and \bea B_f = V_{32}  - i\pi V_{2E}
\Omega_{3E} -  \Omega_{32 }\frac{[-\Delta E_2 + i\pi |V_{2E}|^2 ]}{(E- E_2)} \eea where \bea
E^0_{shift} =  \int \int d r' d r \phi_{3}(r) \Omega_{1}^{*}(r) {\rm Re}[G_E(r',r)] \Omega_{1}
(r')\phi_{3}(r'),\label{E0S} \eea
 \bea V_{32} =  \int \int d r' d r
\phi_{2}(r) V(r){\rm Re}[G_E(r',r)] \Omega_{1}^*(r')\phi_{3}(r') \eea  and  \bea \Delta E_2
=  \int \int d r' d r \phi_{2}(r) V^{*} (r){\rm Re}[G_E(r',r)] V(r')\phi_{2}(r').\eea Here we introduce two new parameters $\beta = (E - E_2 - \Delta E_2)/(\hbar\Gamma_r/2)$ and $ q =(\Omega_{32} + V_{32})/(\pi \Omega_{3E} V_{2E})$.  Now after some algebra we can express (\ref{apn}) in the following form \bea
\tilde{A}_{PA}= \frac{\exp(i\eta_0) (\beta + q ) \pi \Omega_{3 E} }{( \beta + i ) \{\Delta_p + i \hbar (\gamma + \Gamma_{PA})/2\} - \hbar\Gamma_{PA} (q - i)^2 / 2} \label{ap}\eea

\end{document}